# High-pressure structural investigation of several zircon-type orthovanadates

D. Errandonea[1,3], R. Lacomba-Perales[1,2], J. Ruiz-Fuertes[1,2], A. Segura[1,2], S. N. Achary[4], and A. K. Tyagi[4]

[1] Malta Consolider Team, [2]Departamento de Física Aplicada-ICMUV, [3]Fundación General de la Universidad de Valencia, Edificio de Investigación, C/Dr. Moliner 50, 46100 Burjassot (Valencia), Spain

[4] Chemistry Division, Bhabha Atomic Research Centre, Trombay, Mumbai 400085, India

**Abstract:** Room temperature angle-dispersive x-ray diffraction measurements on zircon-type $EuVO_4$, $LuVO_4$, and $ScVO_4$ were performed up to 27 GPa. In the three compounds we found evidence of a pressure-induced structural phase transformation from zircon to a scheelite-type structure. The onset of the transition is near 8 GPa, but the transition is sluggish and the low- and high-pressure phases coexist in a pressure range of about 10 GPa. In $EuVO_4$ and $LuVO_4$ a second transition to a M-fergusonite-type phase was found near 21 GPa. The equations of state for the zircon and scheelite phases are also determined. Among the three studied compounds, we found that $ScVO_4$ is less compressible than $EuVO_4$ and $LuVO_4$, being the most incompressible orthovanadate studied to date. The sequence of structural transitions and compressibilities are discussed in comparison with other zircon-type oxides.

## I. Introduction

Orthovanadates ($AVO_4$, where A is a trivalent element) have recently emerged as promising optical materials for birefringent solid-state laser applications [1, 2]. They can be also employed in a number of applications including their use as cathodoluminescent materials, thermo-phosphors, and scintillators [3]. Most of the orthovanadates crystallize in a tetragonal zircon-type structure (space group: $I4_1/amd$) [3 - 5]; being some of the exceptions triclinic $AlVO_4$ (space group: $P\bar{1}$) [6], tetragonal scheelite-type $BiVO_4$ (space group: $I4_1/a$) [7], monoclinic $InVO_4$ (space group: $C2/m$) [8], and dimorphic $LaVO_4$, which has a zircon-type structure or a monoclinic structure (space group: $P2_1/n$) depending on the mode of preparation [9]. The zircon-type structure consist of isolated $VO_4$ tetrahedra which surround the A atom to form $AO_8$ triangular dodecahedra (bisdisphenoids). The principal structural unit in zircon is a chain of alternating $VO_4$ and $AO_8$ polyhedra extending parallel to the $c$-axis. The chains are joined laterally by edge-sharing $AO_8$ dodecahedra and are responsible for the zircon's prismatic habit and (100) cleavage as well as for its extreme birefringence [10].

Given the technological importance of zircon-type orthovanadates, their electronic and optical properties have been extensively studied [11]. In contrast, their mechanical properties, which are of interest in several areas of materials research, have been studied only for a few of them. Several of the efforts have been dedicated to thermal expansion studies [12, 13] on such zircon-type orthovanadates. In addition, Brillouin-scattering studies have been performed to determine elastic constants [14]. Regarding the behaviour upon compression of zircon-type orthovanadates, x-ray diffraction studies have been performed only for $YVO_4$ [15] and $LuVO_4$ [16]. In the case of $YVO_4$, the low-pressure phase irreversibly transforms to a scheelite-type structure at 8.5 GPa. However, in the case of $LuVO_4$, besides the similar zircon to scheelite phase transition, a second transition to a monoclinic fergusonite-type structure was reported beyond 16 GPa. This result, if confirmed for other vanadates, could be very important in generalizing the predicted zircon-scheelite-fergusonite structural sequence for zircon-type oxides under high-pressure [17]. In order to shed more light on the understanding of the mechanical properties of zircon-type $AVO_4$ compounds, we report structural studies of $EuVO_4$, $LuVO_4$, and $ScVO_4$ up to a pressure of about 27 GPa. The present work contributes to achieve a deeper understanding of pressure effects on the crystal structure of zircon-type oxides of both technological and geophysical importance.

## II. Experimental details

### A. Sample preparation

The $AVO_4$ samples used in the experiments were prepared by solid state reaction of appropriate amounts of pre-dried $A_2O_3$ (A = Eu, Lu, and Sc) (Indian Rare Earth Ltd. 99%) and $V_2O_5$ (Alfa-Aesar 99%). Homogeneous mixtures of the reactants were pelletized and heated at 800°C for 24 h and then cooled to ambient temperature. Further, the pellets were reground and heated again at 1100°C (1000°C) for 24 h for $LuVO_4$ and $ScVO_4$ ($EuVO_4$). The samples obtained were characterized by powder x-ray diffraction data recorded on a Philips X-pert Pro diffractometer using Cu-$K_\alpha$ radiation. Single phase samples of zircon-type structure were confirmed in all. The refined unit-cell parameters for these phases are given in Table I, which are in agreement with the earlier reported values [13].

### B. High-pressure experiments

Angle-dispersive x-ray diffraction (ADXRD) experiments at room temperature (RT) and high pressure with $EuVO_4$ up to 25 GPa, $LuVO_4$ up to 24 GPa, and $ScVO_4$ up to 27 GPa were carried out. Experiments were performed at beamline I15 of the Diamond Light Source using a diamond-anvil cell (DAC) and a monochromatic x-ray beam with a wavelength of 0.61486 Å. Samples were loaded in a 200 μm hole of an inconel gasket in a membrane-type DAC with diamond-culet sizes of 500 μm. Ruby grains were loaded with the sample for pressure determination [18] and silicone oil was used as pressure-transmitting medium [19, 20]. The monochromatic x-ray beam was focused down to 30 × 30 μm$^2$ using Kickpatrick-Baez mirrors. A pinhole placed before the sample position was used as a clean-up aperture for filtering out the tail of the focused x-ray beam. The images were collected using a MAR345 image plate located at 423 mm from the sample. They were integrated and corrected for distortions using FIT2D. The structural analysis was performed using POWDERCELL.

### III. Results

#### A. EuVO$_4$

Figure 1 shows a selection of diffraction patterns of EuVO$_4$ measured at different pressures. A closely similar pressure evolution was observed for the x-ray diffraction patterns of LuVO$_4$ and ScVO$_4$. Considering a zircon-type structure (phase I), all the diffraction patterns observed between ambient pressure and 6.7 GPa can be well indexed. However, at 7.8 GPa we observed the appearance of weak peaks in addition to the peaks of phase I. The intensities of these new peaks gradually increase from 7.8 GPa to 15.1 GPa. At the same time the peaks of phase I gradually lost intensity and fully disappear at 15.1 GPa. This phenomenon is illustrated in the Fig. 1 by a sequence of diffraction patterns collected at 10, 13.6, and 15.1 GPa. These results indicate that a phase-transition takes place in EuVO$_4$. The onset of the transition is at 7.8 GPa, but the transformation is not fully completed up to 15.1 GPa. Within this range of pressure, phase I coexists with the high-pressure phase II. The transition is gradual as shown by the continuous change of peak intensities. From 15.1 to 19 GPa there is no additional changes in the diffraction patterns of EuVO$_4$, but at 20.9 GPa we found a broadening of the peaks and the appearance of additional weak peaks. In particular the peak located near 2θ = 6.3º, which can be seen in Fig. 1 in the diffraction pattern collected at 24.9 GPa. Apparently at 20.9 GPa, a second pressure-induced phase transition takes place from phase II to a phase that we will name as phase III. The phase II-to-phase III transformation is reversible. On pressure release, the pattern recorded at 21.2 GPa corresponds to phase III, while that recorded at 15.1 GPa can be assigned to phase II (see Fig. 1). In contrast, the phase I-to-phase II transition appears to be non-reversible as illustrated by the diffraction pattern measured at 0.4 GPa on pressure release. This fact is in good agreement with the irreversible changes detected in luminescence measurements beyond 5 GPa [21].

In order to characterize the crystalline structure of the high-pressure phases we have taken into account different candidate structures previously observed in compounds related to the orthovanadates. In particular we considered the scheelite ($I4_1/a$) [7], monazite ($P2_1/n$) [4], M-fergusonite ($I2/a$) [22], wolframite ($P2/c$) [22], CrVO$_4$-type (*Cmcm*) [23], LaTaO$_4$-type ($P2_1/n$) [22], high-pressure FeVO$_4$-type (*Pbcn*) [24], and α-MnMoO$_4$ (*C2/m*) [25] structures. After a deep examination of the diffraction patterns assigned to phase II, we found that they can be well indexed considering a scheelite-type structure ($I4_1/a$). At 15.1 GPa we obtain for it the following unit-cell parameters $a$ = 5.045(5) Å and $c$= 11.018(9) Å. The transition from phase I-to-phase II is of first order and involves a volume collapse of approximately 10%. The same transition was previously reported in YVO$_4$ [15] and LuVO$_4$ [16] as well as in other zircon-type oxides (e.g. ZrSiO$_4$) [26].

Regarding the changes observed in the diffraction patterns beyond 19 GPa, they are consistent with the occurrence of a scheelite-to-fergusonite structural transition as previously proposed for LuVO$_4$ [16]. Therefore the M-fergusonite structure is a good candidate for the phase III of EuVO$_4$. We found that the diffraction patterns collected at 20.9, 21.2, and 24.9 GPa can be well indexed considering a M-ferguson ite-type structure. The development of a shoulder in the most intense peak of the scheelite-type phase [(112) reflection around 2θ = 12º], the broadening of the diffraction peaks, and the appearance of a weak peak at low angles (around 2θ = 7º) are typical signatures of the scheelite-to-ferguson ite transition in many $ABO_4$ ternary oxides [17]. Therefore, in spite that the quality of the diffraction patterns collected for phase III do not allow the performance of Rietveld refinements, we can affirm that our experiments provide enough evidence to propose that phase III has a M-fergusonite-type structure (*I2/a*). Considering this structure, the following unit-cell parameters $a$ = 5.00(1) Å, $b$ = 10.91(2) Å, c= 4.95(1) Å, and β = 91.6(1)º are obtained for phase III at 24.9 GPa. According to this result, no noticeable volume change occurs at the scheelite-to-fergusonite transition (see Fig. 4).

From our experiments we extracted the pressure evolution of the unit-cell parameters for phases I and II. The results are summarized in Figure 2. As can be seen in the figure, the compression of the zircon-type structure is non-isotropic, being the *c*-axis the less compressible axes. As a consequence of this, the axial ratio (*c/a*) of phase I gradually increases from 0.880 at ambient pressure to 0.887 at 13.6 GPa. This behaviour is shown in Figure 3. Regarding the unit-cell parameters of the scheelite-type structure, we found that compression is also anisotropic, being the *c*-axis the most compressible axes (as happen in many other scheelites) [17]. In particular, *c/a* decreases nonlinearly from 2.225 at ambient pressure to 2.182 at 19 GPa.

From the pressure dependence of the lattice parameters, the unit-cell volumes of different phases of $EuVO_4$ as a function of pressure were also calculated. The results are summarized in Figure 4. We have analysed the volume changes using a third-order Birch-Murnaghan equation of state (EOS) [27]. The obtained EOS parameters for phase

I are: $V_0$ = 333.2(9) Å$^3$, $B_0$ = 149(6) GPa, and $B_0$'= 5.6(6), being these parameters the zero-pressure volume, bulk modulus, and its pressure derivative, respectively. The bulk modulus of zircon-type EuVO$_4$ is comparable with that of zircon-type LuVO$_4$ and YVO$_4$ [15, 16]. The EOS parameters for phase II are: $V_0$ = 299.4(9) Å$^3$, $B_0$ = 199(9) GPa, and $B_0$'= 4.1(9). The EOS fits for both phases are shown as solid lines in Fig. 4. The bulk modulus of phase II is similar to that reported for the scheelite-type phase in LuVO$_4$ [16], but larger than that reported for the scheelite phase of YVO$_4$ [15]. Empirical models have been developed for predicting the bulk moduli of zircon- and scheelite-structured ABO$_4$ compounds [28]. In particular, the bulk modulus of EuVO$_4$ can be estimated from the charge density of the EuO$_8$ polyhedra using the relation $B_0 = 610 Z_i / d^3$, where $Z_i$ is the cationic formal charge of europium, $d$ is the mean Eu–O distance at ambient pressure (in Å), and $B_0$ is given in GPa [28]. Applying this relation a bulk modulus of 134(25) GPa is estimated for the low-pressure phase of EuVO$_4$ and a bulk modulus of 158(29) GPa is estimated for the scheelite-type phase. These estimations reasonably agree with the values obtained from our experiments and indicate that the scheelite-type phase is less compressible than the zircon-type phase.

**B. LuVO$_4$**

The present results of high-pressure structural studies for LuVO$_4$ qualitatively agree with those previously reported [16] and with our own results on EuVO$_4$. In particular, in our experiments the peaks identified with phase II were found at 8.9 GPa (at 8 GPa in Ref. 16) and phases I and II are found to coexist up to 14.4 GPa. A pure diffraction pattern of phase II is only observed at 16 GPa. As illustrated in Fig. 1 for EuVO$_4$, the transition in LuVO$_4$ is also sluggish, changing continuously the intensities associated to the Bragg peaks of phases I and II. The second phase remains stable up to 21.1 GPa. As proposed by Mittal *et al.* [16], we assigned a scheelite-type structure ($I4_1/a$) to phase II, being the unit-cell parameters at 15.4 GPa: $a$ = 4.875(5) Å and $c$ = 10.674(9) Å. This implies the existence of a large volume collapse of about 13% at the phase I-to-phase II transition. On further compression, at 21.9 GPa we detected identical changes as observed in the diffraction patterns of EuVO$_4$ at 20.9 GPa, indicating the occurrence of a transition from phase II to phase III. No additional changes were found in the diffraction patterns up to 23.6 GPa. As suggested by Mittal *et al.* [16] (second transition at 16 GPa), we found that the diffraction patterns of phase III can be indexed considering a M-fergusonite-type structure ($I2/a$). At 21.9 GPa we determined for this structure $a$ = 4.85(1) Å, $b$ = 10.54(2) Å, $c$ = 4.83(1) Å, and β = 90.2(5)º. This suggests that apparently at the second transition there is no detectable volume change (see Fig. 7). On pressure release from 23.6 GPa, phase II was fully recovered at 15.4 GPa remaining stable at ambient pressure. This is in agreement with the typical non-reversibility of the zircon-scheelite transition and with previous results [16]. The small differences in the transition pressures determined in this work and Ref. 16 can be due to the use of different pressure media and different pressure scales.

The pressure evolution of the unit-cell parameters of phases I and II of LuVO4 are summarized in Figure 5. As in the case of EuVO$_4$, the compression of LuVO$_4$ is anisotropic, being the *a*-axis more compressible in phase I and the *c*-axis more compressible in phase II. The non-isotropic compression of LuVO$_4$ is illustrated in Figure 6. In the zircon-type phase, the $c/a$ ratio increases from 0.887 at ambient pressure to 0.896 at 14.4 GPa. In the scheelite-type phase it decreases from 2.228 at ambient pressure to 2.175 at 21 GPa. From the pressure dependence of the unit-cell parameters, the volume of the different phases of LuVO$_4$ as a function of pressure is calculated. A summary can be found in Figure 7.

We have analysed the volume changes using a third-order Birch-Murnaghan EOS [27] and the obtained EOS parameters for phase I are: $V_0$ = 307.9(9) Å$^3$, $B_0$ = 166(7) GPa, and $B_0$'= 5.6(6). The bulk modulus is comparable with those reported in the literature and the value obtained using the phenomenological model of Ref. 28 as well as from *ab-initio* calculations [16]. Similarly, the obtained EOS parameters for phase II are: $V_0$ = 271.4(9) Å$^3$, $B_0$ = 195(9) GPa, and $B_0$'= 4.9(9). These values are also comparable with previous reported values [16, 28]. The EOS fits for both phases are shown as solid lines in Figure 7. A comparison of different values of the bulk moduli is shown in Table II. Note that again the scheelite phase is less compressible than the zircon phase.

### C. ScVO$_4$

Experiments on ScVO$_4$ also provide evidences of the occurrence of a zircon-to-scheelite transition. In this case the onset of the transition was detected at 8.7 GPa and the low- and high-pressure phases coexist up to 23.4 GPa. At 27.2 GPa, the recorded x-ray diffraction pattern can be completely indexed with a scheelite-type phase. From this pattern, the determined unit-cell parameters for the scheelite structure ($I4_1/a$) of ScVO$_4$ are: $a$ = 4.734(5) and $c$ = 10.374 (9) Å. As in the case of the other two compounds the phase transition here is also irreversible being the scheelite phase recovered at ambient pressure on decompression. In the present case, the volume collapse between the low- and high-pressure phases is around 9 %.

The pressure evolutions for the unit-cell parameters for both phases of ScVO$_4$, obtained from our experiments, are summarized in Figure 8. As observed in the other orthovanadates, in the zircon-type structure of ScVO$_4$, the *c*-axis is less compressible than the *a*-axis. In particular the axial ratio increases from 0.904 at ambient pressure to 0.913 at 23.4 GPa. In the scheelite-type structure the opposite behaviour is observed, the *a*-axis is less compressible than the *c*-axis, decreasing upon compression the axial ratio from 2.228 at ambient pressure to 2.191 at 27.2 GPa. The effects of pressure on the axial ratio are illustrated in Figure 9.

Finally, we have determined the pressure dependence of the volume using a third-order Birch-Murnaghan EOS [27] (see Fig. 10). The fitted EOS parameters for phase I are: $V_0 = 281.1(9)$ Å$^3$, $B_0 = 178(9)$ GPa, and $B_0' = 5.9(9)$. According to this result, ScVO$_4$ is the less compressible zircon-structured vanadate among those studied up to now (see Table II). For phase II we obtained: $V_0 = 256.9(9)$ Å$^3$, $B_0 = 210(12)$ GPa, and $B_0' = 5.3(8)$. As in EuVO$_4$ and LuVO$_4$, the scheelite phase of ScVO$_4$ is also less compressible than the zircon phase.

**IV. Discussion**

A number of structure types of ABO$_4$ compounds with a large difference between the sizes of A and B atoms consist of AO$_8$ dodecahedra and BO$_4$ tetrahedra. These structures include some important mineral structures as zircon (ZrSiO$_4$) and scheelite (CaWO$_4$). It has been shown that these structures are closely and simply related via crystallographic twin operations [10]. In particular, starting with zircon and twinning on (200), (020), and (002) generates the scheelite structure. Because of these symmetry relations the axial ratio of zircon ($c/a \approx 0.9$) is approximately equal to *2a/c* in scheelite (i.e. $c/a \approx 2.2$) as observed in our experiments for EuVO$_4$, LuVO$_4$, and ScVO$_4$. Based upon these crystallographic relations and the correspondences between the zircon (scheelite) and rutile (fluoride) structures [17], the zircon-scheelite transition has been proposed as the most probable high-pressure transformation of zircon-type ABO$_4$ compounds. The existence of such a transition has been confirmed in silicates [26, 30], chromates [31], and phosphates [32]. Exception to this systematic behaviour are TbPO$_4$ [33], which has been proposed to transit from the zircon structure to a monazite-type structure, and CeVO$_4$, which follows the zircon-monazite-scheelite sequence [34]. In the case of the orthovanadates, previous studies reported the zircon-scheelite transition for YVO$_4$ and LuVO$_4$ [15, 16]. Scheelite is also known to be a high-pressure phase of ErVO$_4$ [35]. Our results confirm the occurrence of this transition for LuVO$_4$ and show that EuVO$_4$ and ScVO$_4$ follow the same high-pressure behaviour than most of the zircon-type ABO$_4$ compounds. This conclusion is important since Sc, Y, Lu, Eu, and Er are elements with a quite different population of the 4$f$ sub-shell (e.g. the electronic configuration in Sc is *3d$^1$4s$^2$*, in Eu is *6s$^2$4f$^7$*, and in Lu is *6s$^2$4f$^{14}$5d$^1$*). It was thought that in lanthanide metals like Eu a possible 4$f$ spin-lattice coupling could cause strong anomalies in the high-pressure dependence of AVO$_4$ compounds, as indeed happen with their temperature dependence [36]. In particular, a strong 4$f$ spin-lattice coupling may lead to a cooperative Jahn-Teller transition, lowering the symmetry of the crystal and the symmetry of the lanthanide spins [16]. Also, a pressure-induced *f*-electron delocalisation (which occurs in the lanthanides beyond 10 GPa [37, 38]) could cause a drastic reduction of the atomic bond lengths [39]. However, according to our results, EuVO$_4$ behaves pretty similar to ScVO$_4$ and LuVO$_4$ ruling out the possibility that up to 27 GPa 4$f$ electrons could affect the high-pressure behaviour of rare-earth orthovanadates.

According to the structural-sequence proposed for ABO$_4$ compounds in Ref. 17, the M-fergusonite structure is expected to be the post-scheelite structure for many of them. Previous x-ray diffraction studies reported the zircon-scheelite-fergusonite structural sequence in LuVO$_4$ [16], but did not find any post-scheelite structure in YVO$_4$ up to 26 GPa. However, a broadening of the diffraction peaks was found in YVO$_4$ at 20 GPa and Raman spectroscopy studies reported evidence of the scheelite-to-fergusonite transition in YVO$_4$ beyond 20 GPa [40]. In addition, the softening of the low-frequency T(B$_g$) Raman mode in CaCrO$_4$ and YCrO$_4$ [41, 42] has been used to predict the occurrence of the scheelite-to-fergusonite upon compression in these oxides [40]. In our experiments we confirmed the scheelite-to-fergusonite transition in LuVO$_4$ and also detected it in EuVO$_4$, however we did not find any post-scheelite transition in ScVO$_4$ up to 27.2 GPa. Further high-pressure studies on zircon-type ABO$_4$ compounds are needed to fully understand the structural stability of their high-pressure scheelite phase. It is important to note here that in contrast with the zircon-scheelite transition [26], the scheelite-fergusonite transition is a second-order transition which involves small atomic movements [43]. This fact explains why the second transition is reversible in EuVO$_4$ and LuVO$_4$, but the first transition is not reversible in the three compounds here studied.

Let us now to comment on to the wider pressure range where the scheelite phase of ScVO$_4$ is found to coexist with zircon in comparison with LuVO$_4$ and EuVO$_4$. We think this phenomenon could be related by the smaller compressibility of ScVO$_4$ (see Table II). It seems reasonable to link the described behaviour with the higher strength of the Sc-O bond with respect to the strength of the Lu-O and Eu-O bonds, which correlate with the compressibility (see discussion below), since the zircon-scheelite phase transition mainly involves breaking of A-O bonds [26]. Therefore, in the less compressible compounds one should expect the zircon and scheelite phases to coexist in a wider pressure range.

The pressure evolutions of the cation-oxygen distances have been analyzed for understanding further on this structural transition sequence. The results obtained for EuVO$_4$ are shown in Figure 11. These results are representative of the behaviour observed in the three studied compounds. For them we found that in the zircon phases the V-O distance remains nearly constant within the experimental accuracy, however the Eu-O, Lu-O, and Sc-O distances decrease around 2% from ambient pressure to 10

GPa. A qualitatively similar behaviour occurs for the bond distances in the high-pressure scheelite phases. Additionally, at the phase transition there is no noticeable change in the V-O distances, but the Eu-O, Lu-O, and Sc-O distances are reduced in average about 3%. A similar Y-O bond reduction has been found in YVO$_4$ and also for the Cd-O bond at the high-pressure phase transition of CdV$_2$O$_6$ [44]. The collapse of the A-O bond is related with the large volume contraction observed at the phase transition whereas the different bond-compressibilities explain the anisotropic compression of phases I and II. The zircon structure can be consider as a chain of alternating edge-sharing VO$_4$ tetrahedra and AO$_8$ dodecahedra extending parallel to the *c*-axis, with the chain joined along the *a*-axis by edge-sharing AO$_8$ dodecahedra [10]. The fact that the VO$_4$ tetrahedra behave basically as uncompressible units makes the *c*-axis less compressible than the *a*-axis as observed in our experiments. The same fact causes the anisotropic thermal expansion of the zircon-structured compounds [12, 13]. As a consequence of the symmetry changes between the zircon and the scheelite structure, a rearrangement of the VO$_4$ and AO$_8$ units takes place. This rearrangement provides a more efficient packing, which is consistent with the smaller compressibility of the scheelite phase in comparison with the zircon phase. In addition, in the scheelite structure, VO$_4$ tetrahedra are aligned along the *a*-axis, whereas along the *c*-axis the AO$_8$ dodecahedra are intercalated between the VO$_4$ tetrahedra. Therefore, as the VO$_4$ tetrahedra remain basically undistorted upon compression, in the scheelite structure the *a*-axis is the less compressible axis as found in our experiments.

Based upon the different compressibility of BO$_8$ and AO$_4$ polyhedra in different ABO$_4$ oxides, Hazen and Prewitt [45] found that the bulk modulus of these oxides can be correlated to the compressibility of the AO$_8$ polyhedron. As we mentioned above, most recently [17], it was established the following linear relationship to estimate the bulk modulus in zircon- and scheelite-related structures: $B_0 = 610 Z_i / d^3$, where $Z_i$ is the cationic formal charge of A atom, $d$ is the mean A–O distance at ambient pressure (in Å), and $B_0$ is given in GPa. Given the incompressibility of the VO$_4$ tetrahedra in the orthovanadates, this relationship can be applied to the compounds of interest for this study. According to this, the bulk modulus should increase as the A-O distance at ambient pressure decreases; i.e. those compounds with a larger atomic volume should be the more compressible. This is the behaviour we observed for the low- and high-pressure phases of the studied compounds. Only scheelite EuVO$_4$ apparently slightly deviate from this conduct. A similar tendency is followed by other vanadates as shown in Table II. There, it can be seen, that the empirical relationship proposed in Ref. 28 ($B_0 = 610 Z_i / d^3$) qualitatively agrees with the experimental results, but it tends to underestimate the bulk modulus by about 10%. However, it can be used to extract qualitative conclusions and to make rough estimates of the bulk modulus of unstudied compounds like PrVO$_4$, for which a bulk modulus of 122(24) GPa is predicted. According with this rule, CeVO$_4$ is expected to be the most compressible zircon-type AVO$_4$ compound and ScVO$_4$ the least compressible. Indeed, our results show that ScVO$_4$ has the largest bulk modulus among the compounds already studied. Another conclusion that can be draw is that the scheelite-type phase should be less compressible than the zircon-type phase, as we found in our experiments. If the reduction of the A-O bonds at the transition is around 3%, then the bulk modulus is expected to increase about 9%. This is what we found for EuVO$_4$, LuVO$_4$, and ScVO$_4$. Mittal *et al.* observed the same phenomenon in LuVO$_4$ [16], and other authors found it in several other ABO$_4$ oxides [46, 47]. In contrast with this conclusion, Wang *et al.* [15] found that the bulk modulus of the scheelite phase of YVO$_4$ is only 4% larger than that of the zircon phase.

The above given arguments suggest that probably the bulk modulus of the high-pressure phase of YVO$_4$ needs to be re-determined.

To conclude, we would like to comment on the transition pressures of zircon-type ABO$_4$ compounds. For scheelite-structured ABO$_4$ compounds it was reported that the transition pressure would increase with the ratio $R_{BO_4}/R_A$, where $R_{BO_4}$ and $R_A$ represent the ionic radii of the BO$_4$ units and the A cation [48]. A close inspection to the data available on zircon-structured ABO$_4$ compounds suggests that a similar relationship is not valid for them. According to our and previous results [15, 16, 49], the onset of the zircon-scheelite transition pressure for orthovanadates is always around 7 – 8 GPa, independently of the A cation size. The same can be concluded for the silicates and phosphates, with transition pressures around 20 GPa [30, 32, 47], and the chromates with transition pressures below 6 GPa [31, 41, 42]. In the case of the silicates and phosphates, probably, the compacted SiO$_4$ and PO$_4$ polyhedra make the zircon structure more stable than in other compounds. In the cases of the vanadates and chromates, cooperative interactions between the 3$d$ electrons of the transition metals could make the zircon structure less stable under compression [31]. These facts might explain the different transition pressures found for different families of compounds.

## V. Conclusions

We performed RT ADXRD measurements on zircon-type EuVO$_4$, LuVO$_4$, and ScVO$_4$ up to pressures close to 27 GPa. In the first two compounds, we found the occurrence of two post-zircon phase transitions near 8 GPa and 21 GPa respectively. In ScVO$_4$ we detected only one phase transition at 8.7 GPa. Regarding the crystalline structure of the high-pressure phases we propose a tetragonal scheelite-type and a monoclinic M-fergusonite-type structure. The first transition is sluggish and irreversible while the second transition is reversible. The zircon-scheelite-fergusonite structural sequence is consistent with that deduced from other $ABO_4$ oxides [17]. Regarding possible anomalies related with the occupation of the 4*f*-electron sub-shell in lanthanides such as Eu, we found that $EuVO_4$ follows the same high-pressure behavior that the rest of the orthovanadates. The equation of state for the zircon- and scheelite-type phases has been determined too, finding that $ScVO_4$ is the less compressible vanadate. We also found that the compression of the low- and high-pressure phases is anisotropic. Finally, for both phases we found a differential polyhedral compressibility, behaving the $VO_4$ tetrahedra as rigid units. This fact is related with the anisotropic compressibility of the low- and high-pressure phases.

**Acknowledgments**

Research financed by the Spanish MICINN and the Generalitat Valenciana under Grants No. MAT2007-65990-C03-01, CSD-2007-00045, and GV2008-112. The x-ray diffraction measurements were carried out with the support of the Diamond Light Source at the I15 beamline under proposal No. 683. The authors thank A. Kleppe for technical support during the experiments. R. Lacomba-Perales thanks the support from the MICINN through the "FPU" program.

**References**

[1] W. Ryba-Romanowski, Cryst. Res. Technol. **38**, 225 (2003).

[2] S. Miyazawa, Optoelectron. Rev. **11**, 77 (2003).

[3] D. F. Mullica, E. L. Sappenfield, M. M. Abraham, B. C. Chakoumakos, and L. A. Boatner, Inorg. Chimica Acta **248**, 85 (1996).

[4] A. T. Aldred, Acta Cryst. B **40**, 569 (1984).

[5] B. C. Chakoumakos, M. M. Abraham, and L. A. Boatner, J. Sol. State Chem. **109**, 197 (1994).

[6] U. G. Nielsen, A. Boisen, M. Brorson, C. J. H. Jacobsen, H. J. Jakobsen, and J. Skibsted, Inorg. Chem. **41**, 6432 (2002).

[7] A. W. Sleight, H. Y. Chen, A. Ferreti, and D. E. Cox, Mater. Res. Bull. **14**, 1571 (1979).

[8] M. Touboul, K. Melghit, P. Benard, and D. Loues, J. Sol. State Chem. **118**, 93 (1985).

[9] C. E. Rice and W. R. Robinson, Acta Cryst. B **32**, 2232 (1976).

[10] H. Nyman, B. G. Hyde, and S. Andersson, Acta Cryst. B **40**, 411 (1984).

[11] W. L. Fan, Y. X. Bu, X. Y. Song, S. X. Sun, X. A. Zhan, Cryst. Growth and Design **7**, 2361 (2007) and references therein.

[12] C. V. Reddy, K. S. Murthy, and P. Kistaiah, J. Phys. C: Solid State Phys. **21**, 863 (1988) and references therein.

[13] S. J. Patwe, S. N. Achary, and A. K. Tyagi, Amer. Miner. **94**, 98 (2009).

[14] Y. Hirano, I. Guedes, M. Grimsditch, C. K. Long., N. Wakabayashi, and L. A. Boatner, J. Amer. Ceram. Soc. **85**, 1001 (2002).

[15] X. Wang, I. Loa, K. Syassen, M. Hanfland, and B. Ferrand, Phys. Rev. B **70**, 064109 (2004).

[16] R. Mittal, A. B. Garg, V. Vijayakumar, S. N. Achary, A. K. Tyagi, B. K. Godwal, E. Busetto, A. Lausi, and S. L. Chaplot, J. Phys.: Condens. Matter **20**, 075223 (2008).

[17] D. Errandonea and F. J. Manjon, Progress in Materials Science **53**, 711 (2008).

[18] H. K. Mao, J. Xu, and P. M. Bell, J. Geophys. Res. **91**, 4673 (1986).

[19] Y. Shen, R. S. Kumar, M. Pravica, and M. F. Nicol, Rev. Sci. Instrum. **75**, 4450 (2004).

[20] D. Errandonea, Y. Meng, M. Somayazulu, and D. Häusermann, Physica B **355**, 116 (2005).

[21] G. Chen, R. G. Haire, J. R. Peterson, and M. M. Abraham, J. Phys. Chem. Solids **55**, 313 (1994).

[22] D. Errandonea, J. Pellicer-Porres, F. J. Manjón, A. Segura, Ch. Ferrer-Roca, R. S. Kumar, O. Tschauner, J. López-Solano, P. Rodríguez-Hernández, S. Radescu, A. Mujica, A. Muñoz, and G. Aquilanti, Phys. Rev. B **73**, 224103 (2006).

[23] B. C. Frazer and P. J. Brown, Phys. Rev. **125**, 1283 (1962).

[24] F. Laves, Acta Cryst. **17**, 1476 (1964).

[25] D. Errandonea, M. Somayazulu, and D. Häusermann, phys. stat. sol. (b) **231**, R1 (2002).

[26] M. Marques, J. Contrera-Garcia, M. Florez, and J. M. Recio, J. Phys. Chem. Solids **69**, 2277 (2008).

[27] F. Birch, J. Geophys. Res. **83**, 1257 (1978).

[28] D. Errandonea, J. Pellicer-Porres, F. J. Manjón, A. Segura, Ch. Ferrer-Roca, R. S. Kumar, O. Tschauner, P. Rodríguez-Hernández, J. López-Solano, S. Radescu, A. Mujica, A. Muñoz, and G. Aquilanti, Phys. Rev. B **72**, 174106 (2005).

[29] R. M. Hazen, Science **216**, 991 (1982).

[30] B. Manoun, R. T. Downs, S. K. Saxena, Amer. Mineral. **91**, 1888 (2006).

[31] Y. W. Long, L. X. Yang, Y. Yu, F. Y. Li, Y. X. Lu, R. C. Yu, L. Liu, and C. Q. Jin, J. Appl. Phys. **103**, 093542 (2008) and references therein.

[32] F. X. Zhang, M. Lang, R. C. Ewing, J. Lian, Z. W. Wang, J. Hu, and L. A. Boatner, J. Solid State Chem. **181**, 2633 (2008).

[33] A. Tatsi, E. Stavrou, Y. C. Boulmetis, A. G. Kontos, Y. S. Raptis, and C. Raptis, J. Phys.: Condens. Matter **20**, 425216 (2008).

[34] K. J. Range, H. Meister, and U. Klemt, Z. Naturf. B **45**, 598 (1990).

[35] K. J. Range and H. Meister, Acta Cryst. C **46**, 1093 (1990).

[36] S. Skanthakumar, C. K. Loong, L. Soderholm, J. W. Richardson, M. M. Abraham Jr., and L. A. Boatner, Phys. Rev. B **51**, 5644 (1995).

[37] D. Errandonea, R. Boehler, and M. Ross, Phys. Rev. Letters **85**, 3444 (2000).

[38] A. Svane, W. Temmerman, and Z. Szotek, Phys. Rev. B **59**, 7888 (1999).

[39] H. W. Sheng, H. Z. Liu, Y. Q. Cheng, J. Wen, P. L. Lee, W. K. Luo, S. D. Shastri, and E. Ma, Nature Materials **6**, 192 (2007).

[40] D. Errandonea and F. J. Manjon, Materials Research Bulletin **44**, 807 (2009).

[41] Y.W. Long, W.W. Zhang, L.X. Yang, Y. Yu, R.C. Yu, S. Ding, Y.L. Liu, C.Q. Jin, Appl. Phys. Lett. **87**, 181901 (2005).

[42] Y.W. Long, L.X. Yang, Y. Yu, F.Y. Li, R.C. Yu, S. Ding, Y.L. Liu, C.Q. Jin, Phys. Rev. B **74**, 054110 (2006).

[43] D. Errandonea, EPL **77**, 56001 (2007).

[44] A. A. Belik, A. V. Mironov, R. V. Shpanchenko, and E. T. Muromachi, Acta Cryst. C **63**, i37 (2007).

[45] R. M. Hazen and C. T. Prewitt, Amer. Mineral. **62**, 309 (1997).

[46] V. Panchal, N. Garg, S. N. Achary, A. K. Tyagi, and S. M. Sharma, J. Phys.: Condens. Matter **18**, 8241 (2006).

[47] H. P. Scott, Q. Williams, and E. Knittle, Phys. Rev. Lett. **88**, 015506 (2002).

[48] D. Errandonea, F. J. Manjón, M. Somayazulu, and D. Häusermann, J. Solid State Cehm. **177**, 1087 (2004).

[49] S. J. Duclos, A. Jayaraman, G. P. Espinosa, A. S. Cooper, and R. G. Maines, J. Phys. Chem. Solids **50**, 769 (1989).

**Table I:** Unit-cell parameters and atomic coordinates of EuVO$_4$, LuVO$_4$, and ScVO$_4$ at ambient conditions. The three compounds crystallize in the zircon structure (space group: $I4_1/amd$) being the A atoms located at the Wyckoff position 4a (0,3/4,1/8), the V atoms at 4b (0,1/4,3/8), and the O atoms at 16h (0,$u$,$v$).

| Compound | $a$ [Å] | $c$ [Å] | Atomic coordinates |
|---|---|---|---|
| EuVO$_4$ | 7.2357(1) | 6.3657(1) | $u$ = 0.4271(9) |
|  |  |  | $v$ = 0.2119(9) |
| LuVO$_4$ | 7.0230(1) | 6.2305(1) | $u$ = 0.4300(11) |
|  |  |  | $v$ = 0.2064(10) |
| ScVO$_4$ | 6.7805(2) | 6.1346(3) | $u$ = 0.4409(9) |
|  |  |  | $v$ = 0.1972(10) |

**Table II:** Ambient-pressure volume and bulk modulus of different scheelite and zircon type orthovanadates. The $B_0$ values obtained from the present data are compared with previous estimations and with the values obtained following the empirical model proposed in Ref. 28.

| Compound | Structure | $V_0$ [Å$^3$] | $B_0$ [GPa] This work | $B_0$ [GPa] Experiments | $B_0$ [GPa] Ab initio | $B_0$ [GPa] Empirical model |
|---|---|---|---|---|---|---|
| EuVO$_4$ | Zircon | 333.4 | 149(6) | | | 134(26) |
| DyVO$_4$ | Zircon | 325.9 | | 140(5)[a] | | 135(26) |
| TbVO$_4$ | Zircon | 322.2 | | 129(5)[a] | | 137(26) |
| HoVO$_4$ | Zircon | 319.1 | | 142(9)[a] | | 138(26) |
| YVO$_4$ | Zircon | 318.7 | | 130(3)[b] | | 138(26) |
| ErVO$_4$ | Zircon | 315.9 | | 136(9)[a] | | 140(27) |
| LuVO$_4$ | Zircon | 307.7 | 166(7) | 147[c] | 166[c] | 145(28) |
| ScVO$_4$ | Zircon | 282.2 | 178(9) | | | 162(30) |
| BiVO$_4$ | Scheelite | 311.2 | | 150(5)[d] | | 142(27) |
| EuVO$_4$ | Scheelite | 299.4 | 199(9) | | | 158(29) |
| YVO$_4$ | Scheelite | 284.5 | | 138(9)[b] | | 160(29) |
| LuVO$_4$ | Scheelite | 271.4 | 195(9) | 194[c] | 173[c] | 166(30) |
| ScVO$_4$ | Scheelite | 256.9 | 210(12) | | | 185(35) |

[a] Estimated from the elastic constants reported in Ref. 14. [b] Ref. 15. [c] Ref. 16. [d] Ref. 29.

**Figure Captions**

**Figure 1:** Selection of x-ray diffraction patterns measured in EuVO4 at different pressures. Pressures are indicated in the figure. (r) denotes those patterns collected on pressure release. Miller indices of the zircon phase at 0.2 GPa are provided for clarity.

**Figure 2:** Pressure evolution of the unit-cell parameters of the zircon-type and scheelite-type phases of $EuVO_4$. To facilitate the comparison for the high-pressure phase we plotted *c/2* instead of *c*. Symbols: experiments. Lines: quadratic fit.

**Figure 3:** Pressure dependence of the axial ratio in $EuVO_4$. Symbols: experiments. Lines: quadratic fit.

**Figure 4:** Pressure-volume relation in $EuVO_4$. Symbols: experiments. Lines: EOS fit (extrapolated beyond 19 GPa). A data point of phase III is shown (triangle) to illustrate that apparently there is no volume change at the scheelite-fergusonite transition.

**Figure 5:** Pressure evolution of the unit-cell parameters of the zircon-type and scheelite-type phases of $LuVO_4$. To facilitate the comparison for the high-pressure phase we plotted *c/2* instead of *c*. Symbols: experiments. Lines: quadratic fit.

**Figure 6:** Pressure dependence of the axial ratio in $LuVO_4$. Symbols: experiments. Lines: quadratic fit.

**Figure 7:** Pressure-volume relation in $LuVO_4$. Symbols: experiments. Lines: EOS fit (extrapolated beyond 21 GPa). A data point of phase III is shown (triangle) to illustrate that apparently there is no volume change at the scheelite-fergusonite transition.

**Figure 8:** Pressure evolution of the unit-cell parameters of the zircon-type and scheelite-type phases of $ScVO_4$. To facilitate the comparison for the high-pressure phase we plotted *c/2* instead of *c*. Symbols: experiments. Lines: quadratic fit.

**Figure 9:** Pressure dependence of the axial ratio in ScVO$_4$. Symbols: experiments. Lines: quadratic fit.

**Figure 10:** Pressure-volume relation in ScVO$_4$. Symbols: experiments. Lines: EOS fit.

**Figure 11:** Pressure dependence of the interatomic distances in EuVO$_4$. Error bars are of similar size than symbols. The lines are just a guide to the eye.

**Figure 1**

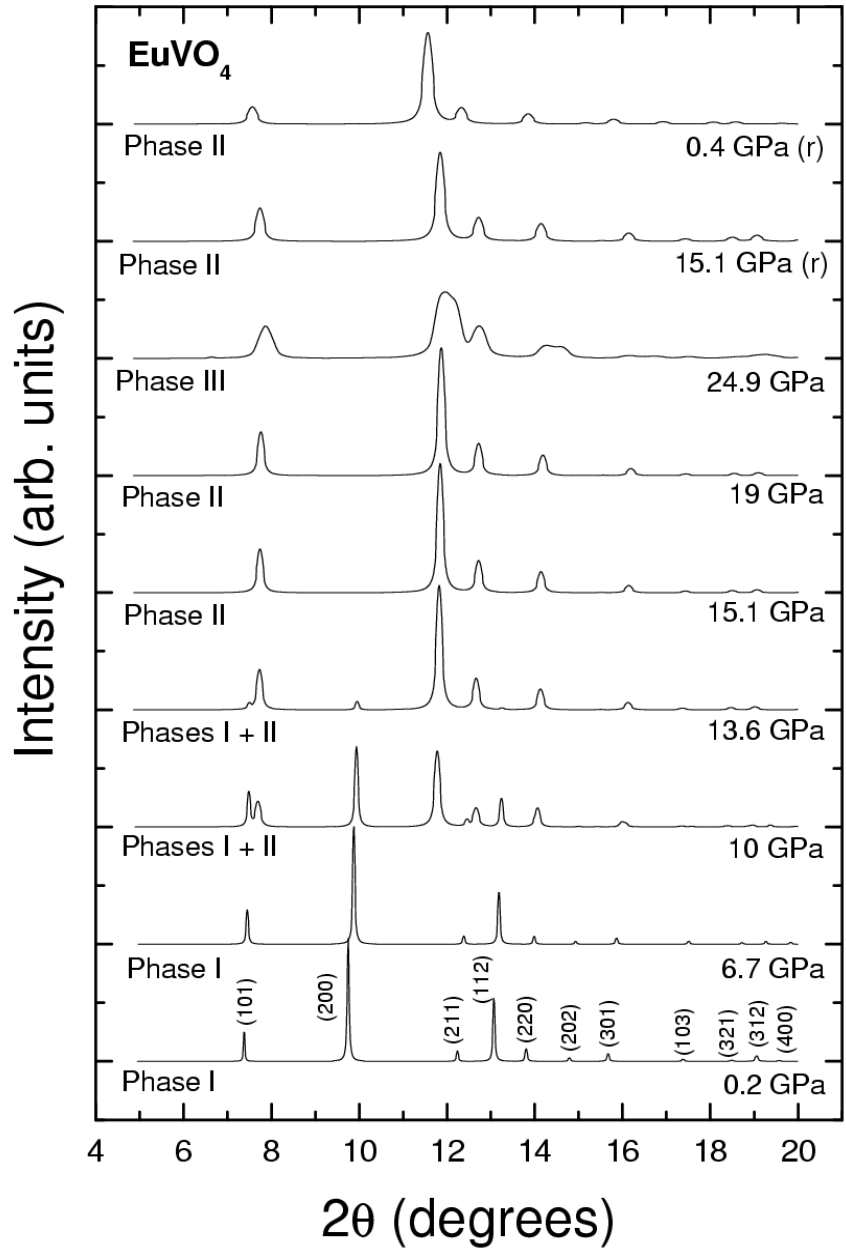

**Figure 2**

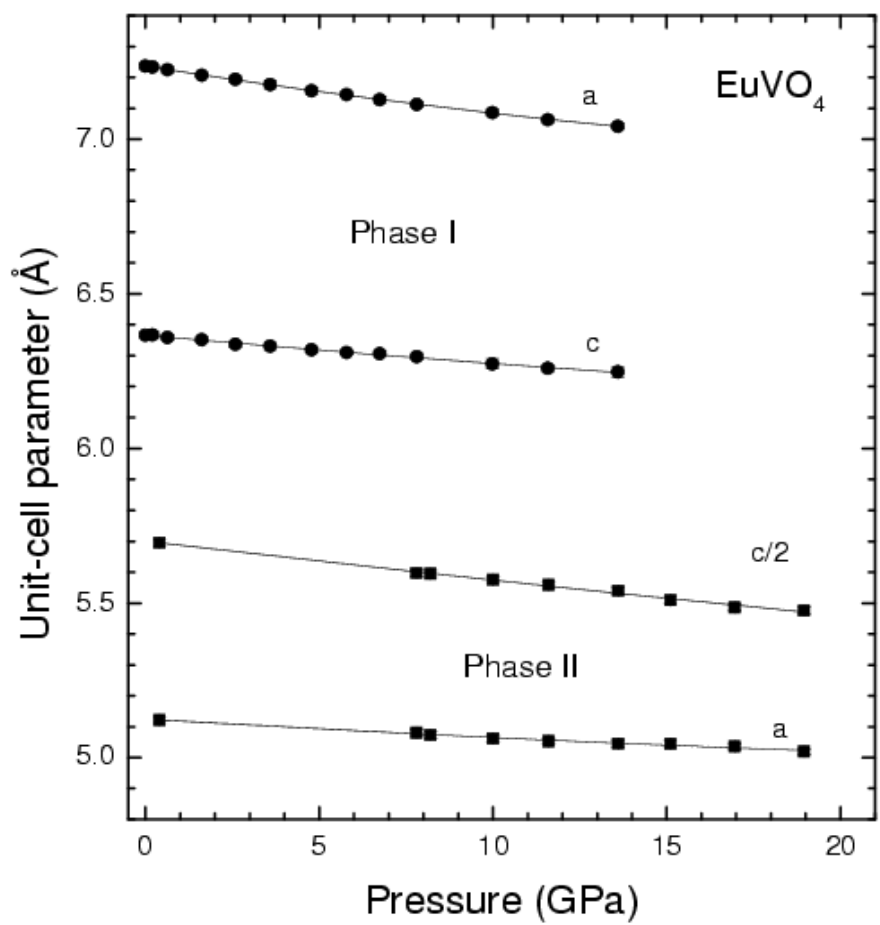

**Figure 3**

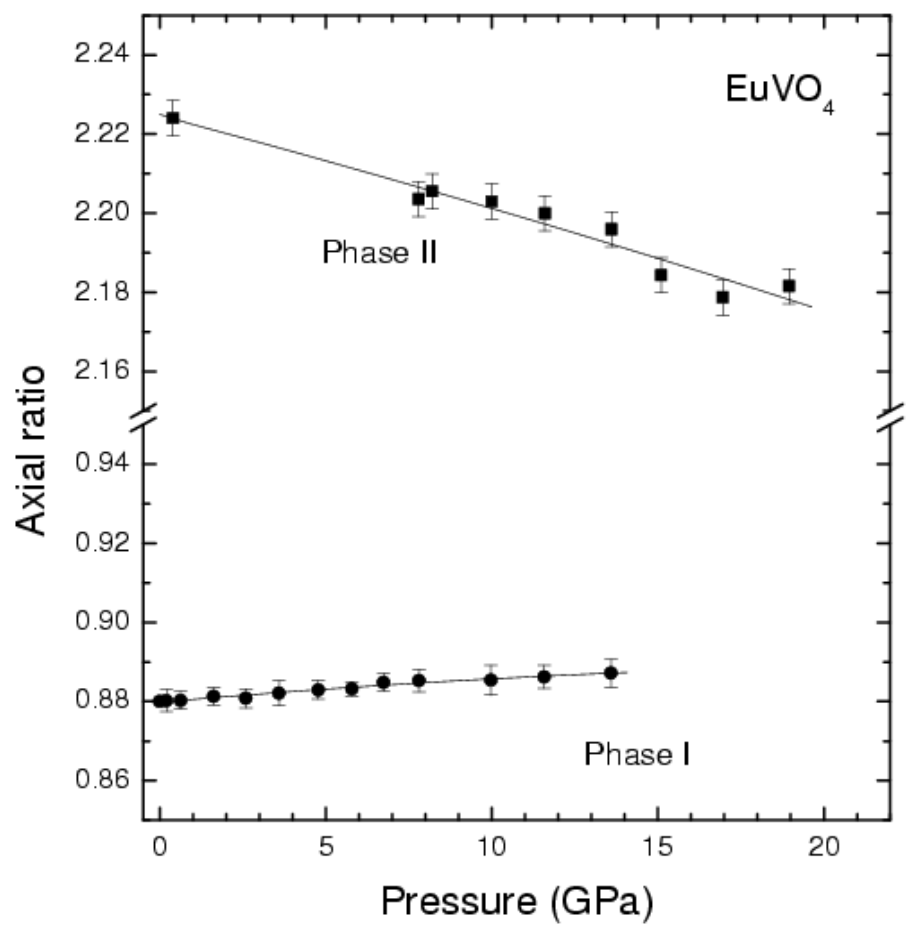

**Figure 4**

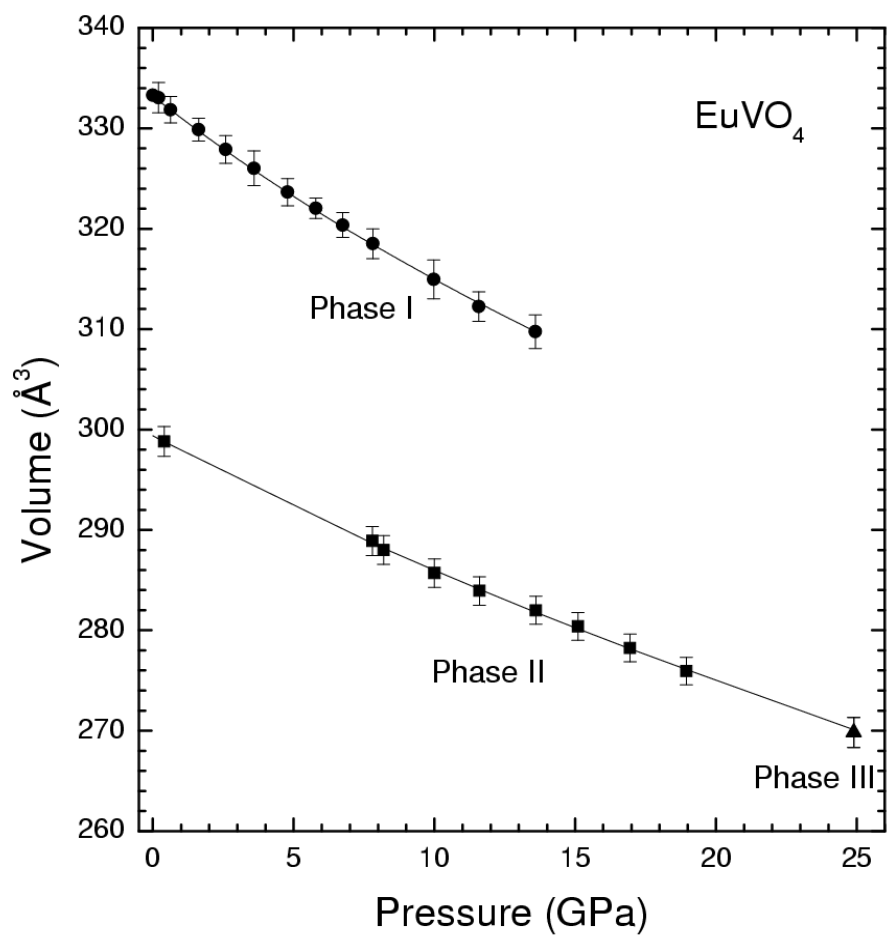

**Figure 5**

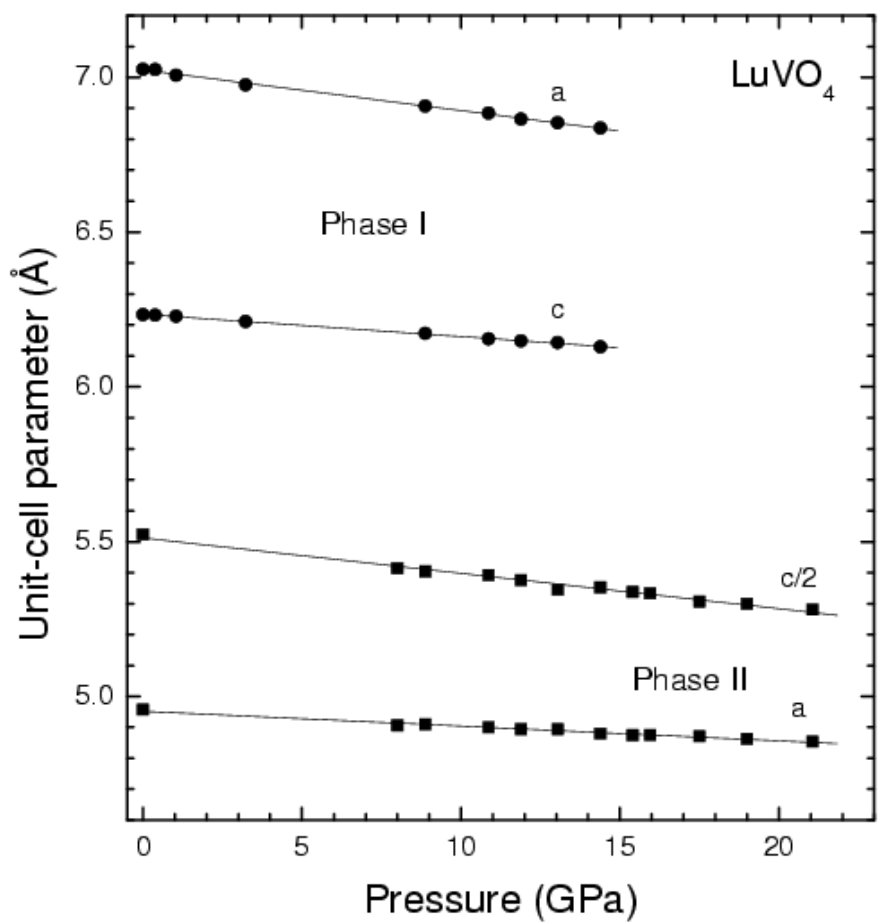

**Figure 6**

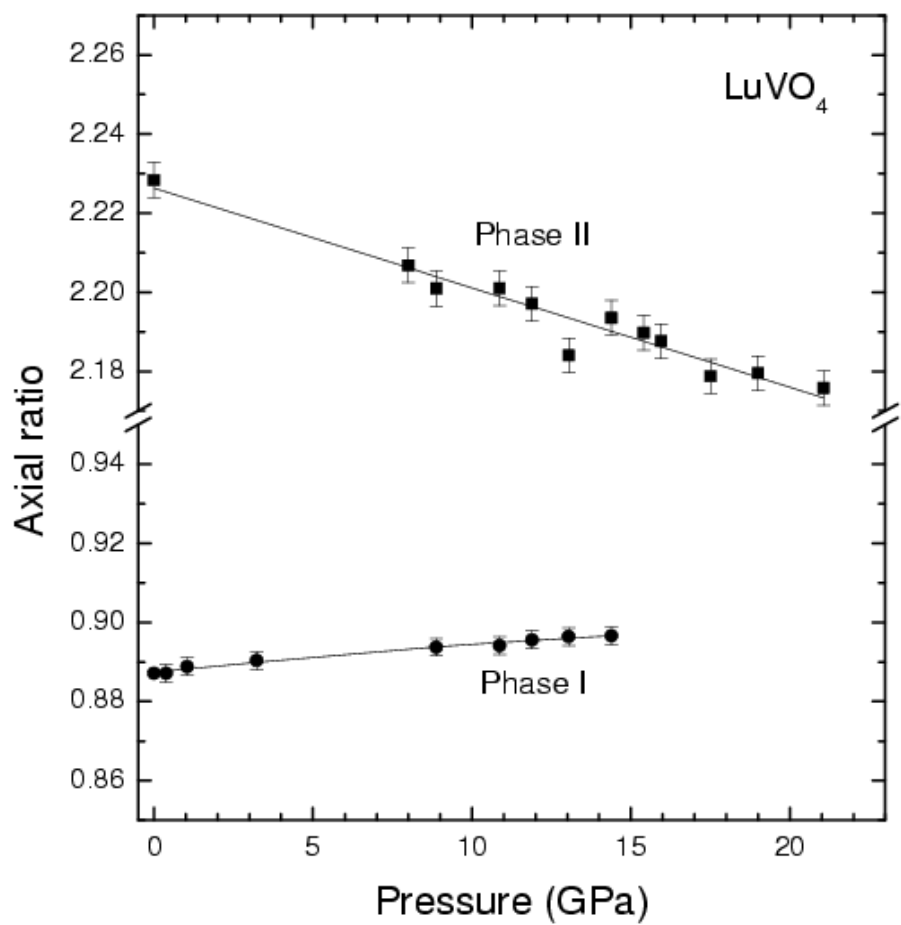

**Figure 7**

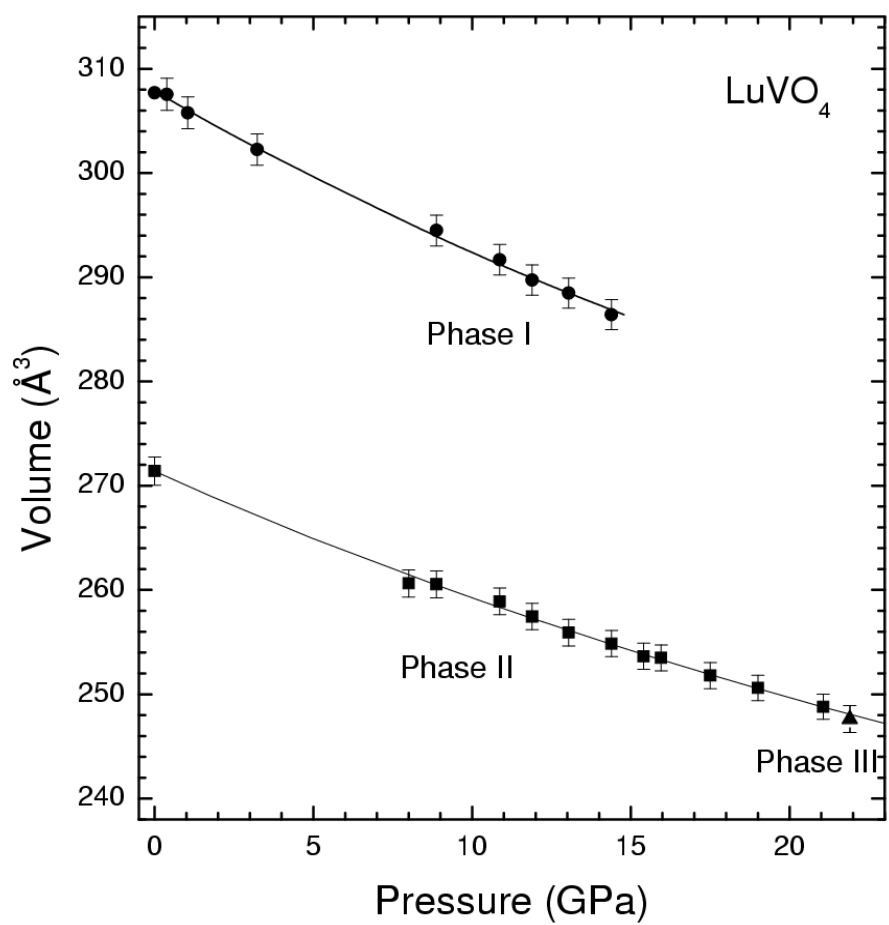

**Figure 8**

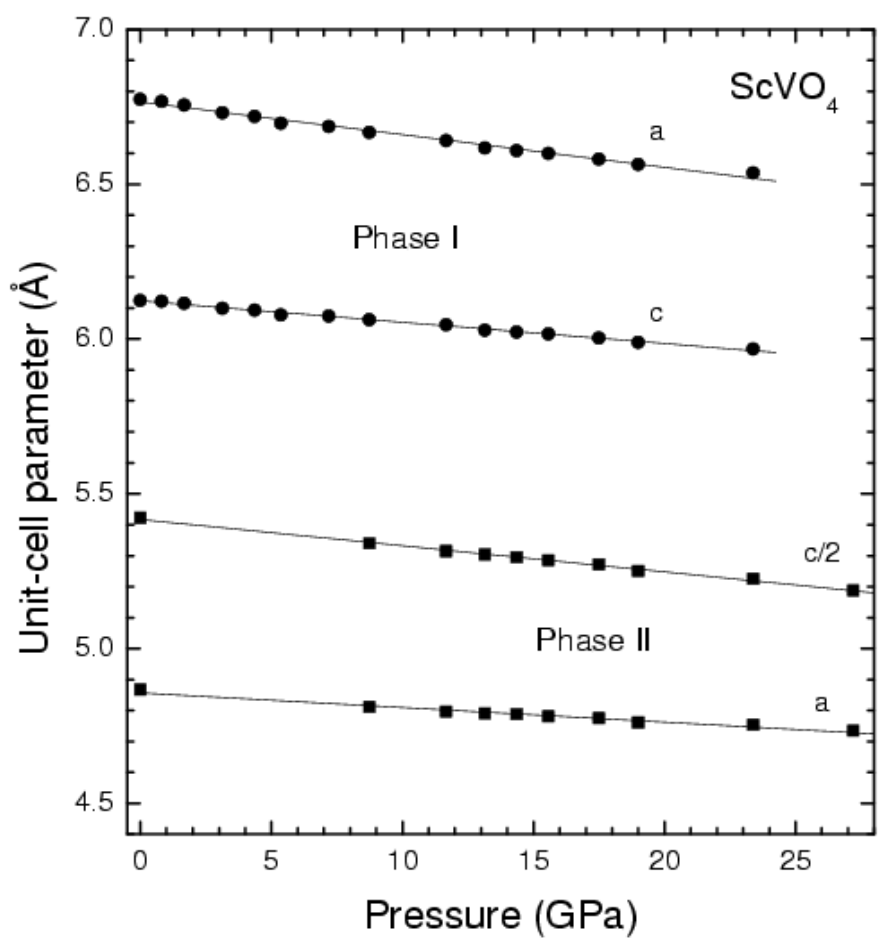

**Figure 9**

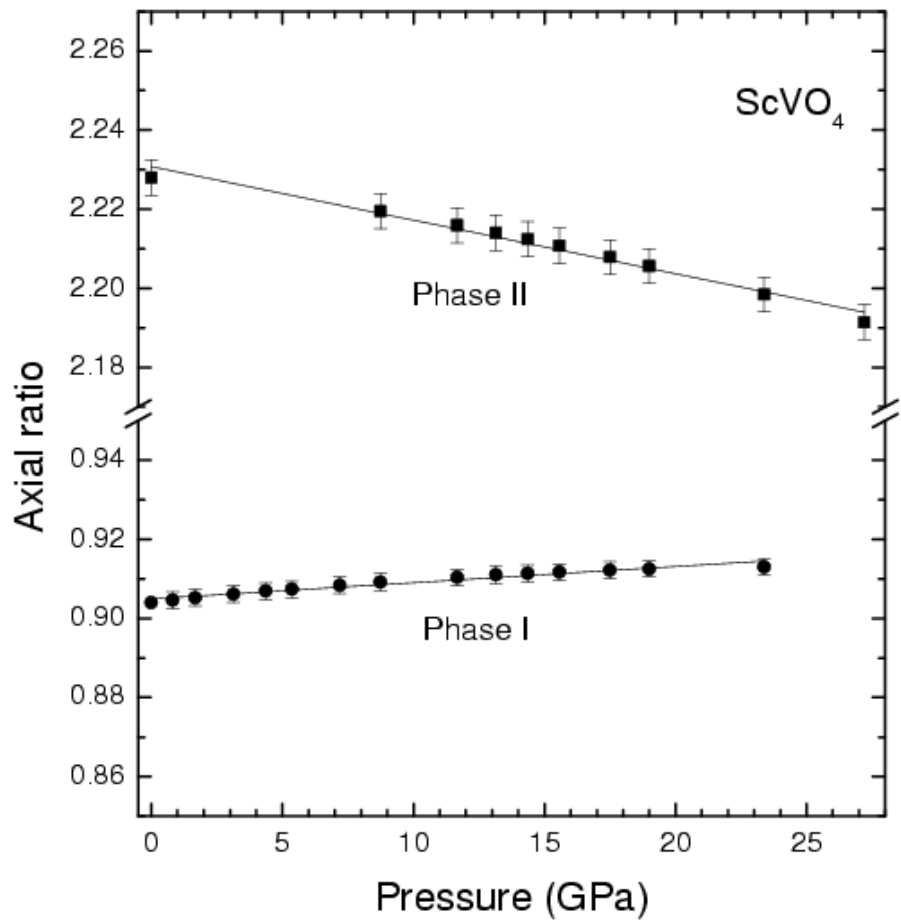

**Figure 10**

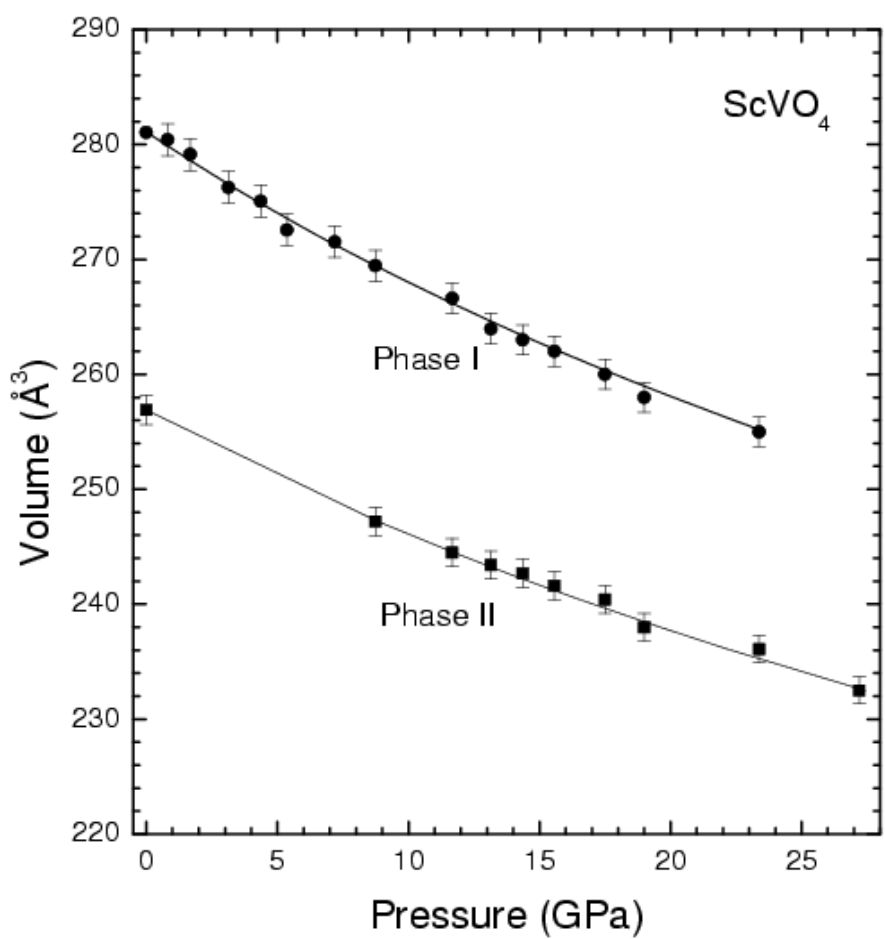

**Figure 11**

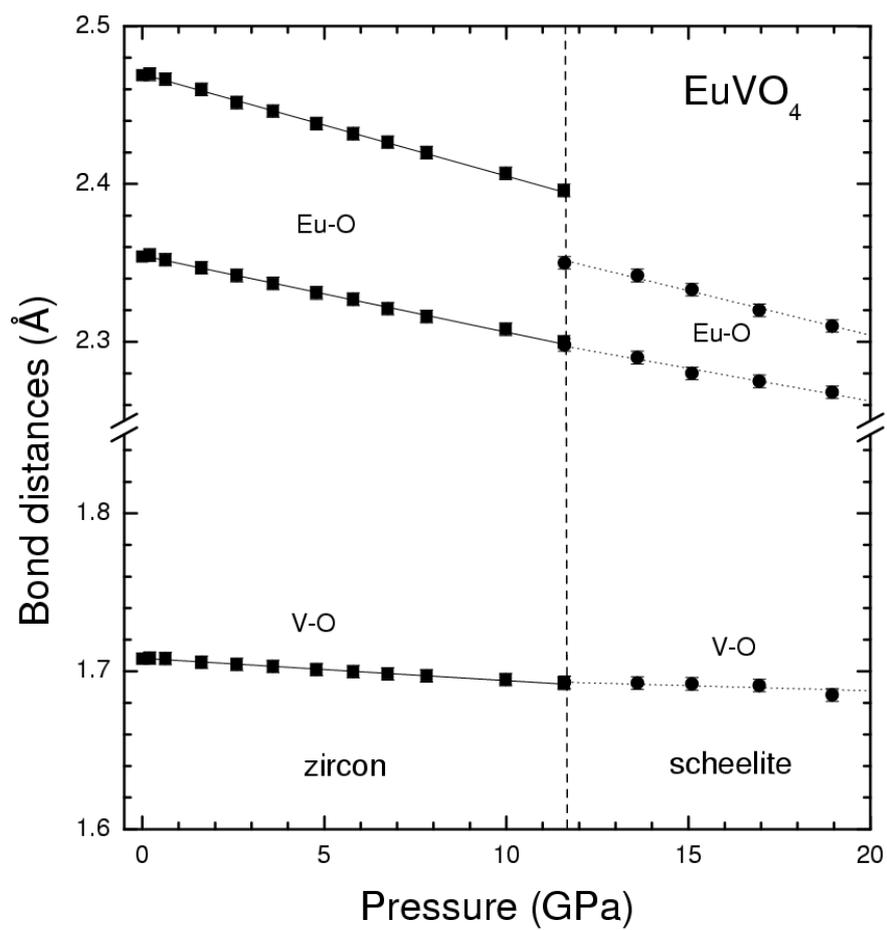